\documentclass[aps,pre,superscriptaddress,showkeys,showpacs]{revtex4}
\usepackage{graphicx,amsmath,amssymb,amsfonts,amstext,bm}

\begin{document}

\title{Wetting in the presence of the electric field:\\
       The classical density functional theory study for a model system}

\author{Vasyl~Myhal}
\affiliation{Ivan Franko National University of L'viv,
              Department for Theoretical Physics,
              Drahomanov Street 12, 79005 L'viv, Ukraine}

\author{Oleg~Derzhko}
\affiliation{Institute for Condensed Matter Physics,
              National Academy of Sciences of Ukraine,
              Svientsitskii Street 1, 79011 L'viv, Ukraine}
\affiliation{Ivan Franko National University of L'viv,
              Department for Theoretical Physics,
              Drahomanov Street 12, 79005 L'viv, Ukraine}
\affiliation{Abdus Salam International Centre for Theoretical Physics,
              Strada Costiera 11, 34011 Trieste, Italy}

\date{\today}

\begin{abstract}
We discuss the effect of an external electric field on the wetting of a solid surface by liquid.
To this end, 
we use a model of the two-level-atom fluid
for which the changes in interatomic interactions due to the presence of the field 
can be found using quantum-mechanical perturbation theory.
Constructing the grand potential functional,
we perform the standard calculations of Young's equilibrium contact angle.
The switching on of the electric field $\vert{\bf {E}}\vert > 0$
may increase noticeably the contact angle $\theta$.
\end{abstract}



\maketitle


\section{Introductory remarks}

Wetting of solid surfaces by liquids is important both from fundamental and practical points of view \cite{gennes,armr,bonn,saam}.
Liquid wets a solid surface, 
if the Young equilibrium contact angle $\theta$
(i.e., the angle between the surface of the liquid and the outline of the contact solid surface at thermodynamic equilibrium)
vanishes, i.e., $\theta = 0^\circ$ (complete wetting).
The surface is nonwetted (partial wetting) for any $0^\circ<\theta< 180^\circ$ 
and it is completely dry for $\theta = 180^\circ$.
The contact angle may vary under the change of external parameters.
If $\theta$ while increasing crosses the value $90^\circ$,
the change from hydrophilicity ($\theta < 90^\circ$) to hydrophobicity ($\theta > 90^\circ$) occurs.

An interesting problem in the theory of inhomogeneous fluids
is to examine a dependence of $\theta$ on external parameters starting from a microscopic picture
within the frames of which one can follow how external influences modify interparticle interactions.
The classical density functional method \cite{evans,oxtoby-review,derzhko-myhal-lectures,loewen,tarazona,roth,lutsko} provides such a possibility
since it allows to calculate the properties of a nonuniform fluid
on the basis of interparticle interactions.

It is well known that an external electric field is a simple and effective way to change wetting properties.
The most drastic changes in the presence of the electric field occur for ionic or polar liquids
(see, for example, Refs.~\onlinecite{kang,bier,rui,bateni}).
The shape and stability of droplets in the electric field, contact angle phenomena in the electric field, 
as well as electrocapilarity
have been of interest for a long time
and receive renewed attention because of electrowetting 
(see, e.g., review papers~\cite{mugele,vancauwenberghe} and references therein).
In electrowetting,
one is generically dealing with droplets of partially wetting conductive liquids (electrolytes) on planar solid substrates
and the applied voltage changes the contact angle.
However, even in the case of noble liquids 
the electric field can affect the macroscopic properties,
and in particular the wetting properties,
via coupling to the transition electric dipole moment of atoms.

In the present paper,
we intend to follow starting from the microscopic level
how an external electric field affects the Young equilibrium contact angle for a fluid of atoms.
To this end, we consider a simple model of two-level-atom fluid
in which the interatomic interactions are changed because of the presence of the field.
Furthermore,
within the frames of the classical density functional theory approach
we calculate the contact angle $\theta$ which depends on the value of the electric field strength $\vert{\bf{E}}\vert$.
We show that while the value of the electric field strength increases, 
the contact angle may increase and cross $90^\circ$.
In other words,
an increase of the field may lead to hydrophobicity.

\begin{figure}
\center{\includegraphics[width=85mm]{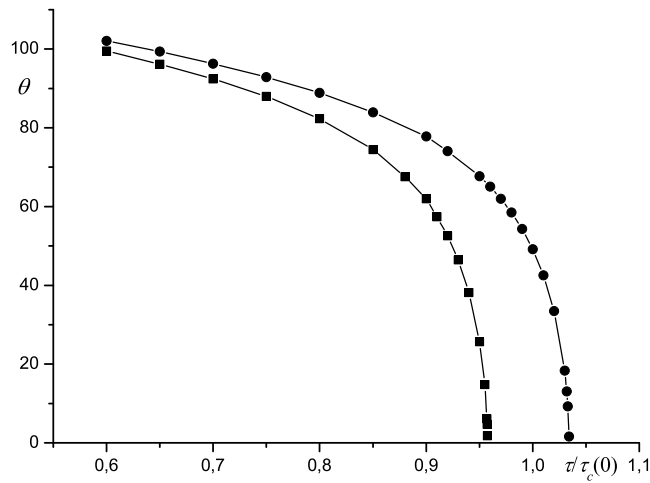}}
\caption[]{Dependence of the contact angle $\theta$ on the temperature $\tau /\tau_c(0)$.
$\tau_c(0)$ denotes the dimensionless critical temperature at ${\bf{E}}=0$,
i.e., $\tau_c(0)=T_c({\bf{E}}=0)/(E_1-E_0)$.
The lower curve (with squares) corresponds to ${\cal{E}}=0$, 
the upper curve (with circles) corresponds to ${\cal{E}}=0.2$;
${\cal{E}}=\vert{\bf{E}}\vert r_0^3/\vert{\bf{p}}\vert$ is the dimensionless value of the electric field strength.
For further explanations see the main text.}
\label{fig01}
\end{figure}

\begin{figure}
\center{\includegraphics[width=85mm]{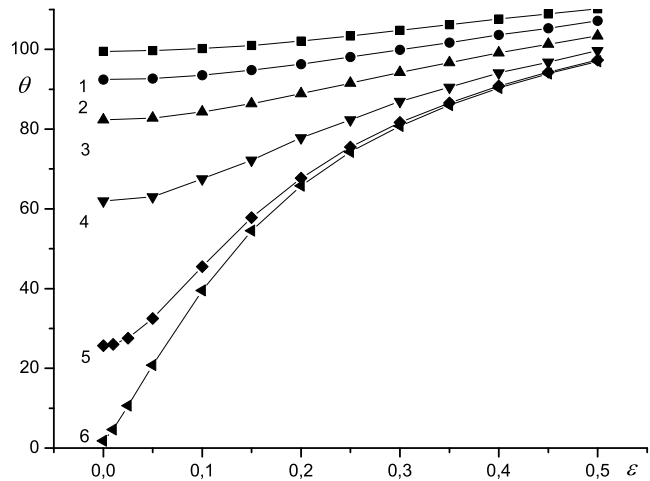}}
\caption[]{Dependence of the contact angle $\theta$ on ${\cal{E}}=\vert{\bf{E}}\vert r_0^3/\vert{\bf{p}}\vert$
at different temperatures $\tau=T/(E_1-E_0)$:
$0.6 \tau_c(0)\approx 0.001\,179$ (curve 1 with squares),
$0.7 \tau_c(0)\approx 0.001\,376$ (curve 2 with circles),
$0.8 \tau_c(0)\approx 0.001\,572$ (curve 3 with up-triangles),
$0.9 \tau_c(0)\approx 0.001\,769$ (curve 4 with down-triangles),
$0.95\tau_c(0)\approx 0.001\,867$ (curve 5 with diamonds),
and
$0.957\,4\tau_c(0)\approx 0.001\,881$ (curve 6 with left-triangles).
For further explanations see the main text.}
\label{fig02}
\end{figure}

The outline of the paper is as follows. 
First, we justify the choice of a grand potential functional
which depends on an external electric field.
Then we report some results for the bulk properties of the system, 
as well as for the density profiles for two-phase cases:
liquid -- vapor, 
substrate (solid wall) -- liquid,
and
substrate -- vapor. 
Knowing the grand potential allows us to find the surface tensions, 
and then, 
via the Young equation,
to obtain the required contact angle $\theta$. 
Our main results are shown in Figs.~\ref{fig01} and \ref{fig02}. 
From these plots one can see 
that an increase of the value of the electric field strength $\vert{\bf {E}}\vert$
increases the wetting temperature $T_w$
(i.e., the temperature $T_w$ for which the contact angle $\theta$ vanishes),
see Fig.~\ref{fig01}, 
increases the contact angle $\theta$ at fixed temperature,
see Fig.~\ref{fig02},
may replace wetting by partial wetting, 
see Figs.~\ref{fig01}, \ref{fig02},
and 
may lead to a changeover from hydrophilicity to hydrophobicity,
see Figs.~\ref{fig01}, \ref{fig02}.

\section{Interatomic interactions and the grand potential functional}

In order to follow how the electric field affects the contact angle 
we have to begin with writing down the interaction energy of neutral atoms with a time-independent spatially uniform electric field.
This might be a puzzle since the atoms have no permanent electric dipole moment. 
Therefore, 
we start from the first principles to show how the electric field modifies the interatomic interactions.
To be free of secondary complications,
we shall consider a simple model of a two-level-atom fluid.
We assume that the energy of excitation of the atom is $E_1-E_0$,
the atom does not have the electric dipole moment in the ground state or in the excited state,
and the transition electric dipole moment between the ground and excited states is ${\bf {p}}$.
We are interested in how the electric field ${\bf {E}}$ modifies the long-range interatomic interactions,
while the short-range interactions are described by introducing the atom radius $r_0=\sigma/2$.
After switching on the electric field $\vert{\bf {E}}\vert > 0$,
one can calculate 
within the framework of the quantum-mechanical perturbation theory with respect to the interaction with the field
the second-order results for the energy 
of a single atom, $E_{N=1}$,
or 
of a group of two atoms at (a sufficiently large) distance $R=\vert {\bf{R}}_{12}\vert$, $E_{N=2}$,
see Appendix and Ref.~\onlinecite{derzhko-myhal-zhfd}.
We find
\begin{eqnarray}
\label{01}
E_{N=1}=E_0-\frac{\gamma_1^2}{4}\left(E_1-E_0\right)+\ldots,
\nonumber\\
E_{N=2}
=
2E_0
-\left(\frac{\alpha_{12}^2}{2}+\frac{\gamma_1^2+\gamma_2^2}{4}\left(1+\frac{3\alpha_{12}^2}{2}\right)-\gamma_1\gamma_2\alpha_{12}\right)
\left(E_1-E_0\right)+\ldots,
\nonumber\\
\gamma_i(E_1-E_0)=2\vert{\bf{p}}\vert \vert{\bf{E}}\vert \chi_i,
\;\;\;
\alpha_{12}(E_1-E_0)=\frac{\vert{\bf{p}}\vert^2}{R^3}\Phi_{12},
\end{eqnarray}
where $\chi_i$ and $\Phi_{12}$ are well known functions which depend 
on ${\bf{p}}_i/\vert {\bf{p}}\vert$ and ${\bf{E}}/\vert {\bf{E}}\vert$
or
on ${\bf{p}}_1/\vert {\bf{p}}\vert$, ${\bf{p}}_2/\vert {\bf{p}}\vert$, and ${\bf{R}}_{12}/\vert {\bf{R}}_{12}\vert$,
see Appendix and Ref.~\onlinecite{derzhko-myhal-zhfd}.
$E_{N=1}$ and $E_{N=2}$ given in Eq.~(\ref{01})
are the only results one must know in order to find the second virial coefficient of the fluid.
The statistical-mechanical average contains also the averages over orientations of ${\bf{p}}_i$
(and therefore no preferential direction created by the field is expected).
Bearing in mind that we are interested in the lowest term in $\vert{\bf {E}}\vert$ only,
the orientational averages can be done using a cumulant expansion.
After straightforward but cumbersome calculations
(for details see Appendix)
we find the second virial coefficient of the two-level-atom fluid \cite{derzhko-myhal-zhfd}:
\begin{eqnarray}
\label{02}
B_2(T, \vert{\bf {E}}\vert)
=
4v-2\pi\int_{\sigma}^{\infty}{\rm{d}}R R^2
\left(
\exp\left(\frac{3a(\vert{\bf{E}}\vert)\sigma^3}{2\pi TR^6}\right) - 1
\right)
\approx
4v-\frac{a(\vert{\bf {E}}\vert)}{T},
\nonumber\\
a(\vert{\bf {E}}\vert)
=
\frac{2\pi}{9}
\left(
1+\frac{2\vert{\bf{p}}\vert^2 \vert{\bf{E}}\vert^2}{(E_1-E_0)^2}
\right)
\frac{\vert{\bf{p}}\vert^4}{(E_1-E_0)\sigma^3}
=
\frac{v (E_1-E_0) \aleph^2}{48}  \left(1+2\aleph^2{\cal{E}}^2\right).
\end{eqnarray}
Here
$v=\pi\sigma^3/6$,
$\aleph = \vert {\rm {\bf p}}\vert ^2 / (r_0^3(E_1-E_0))$
is the dimensionless parameter which characterizes the two-level atom
(in what follows we set $\aleph = 1$ for convenience),
${\cal{E}} =\vert{\bf{E}}\vert r_0^3/\vert{\bf{p}}\vert$
is the dimensionless value of the electric field strength.
For $\vert{\bf {E}}\vert=0$
one immediately recognizes in Eq.~(\ref{02})
the contribution of van der Waals interactions to the second virial coefficient.
For $\vert{\bf {E}}\vert > 0$ the interaction constant of van der Waals interactions increases
in accordance with the rescaling $a({\bf{E}}=0)\to a(\vert{\bf {E}}\vert)= a({\bf{E}}=0)(1+2\aleph^2{\cal{E}}^2)$.

Equation~(\ref{02}) allows us to construct 
an extrapolated equation of state which already contains the liquid-vapor phase transition,
and to find the corresponding Helmholtz free energy and the grand potential,
as well as to extend the latter findings to a nonuniform case,
see reviews \cite{evans,oxtoby-review,derzhko-myhal-lectures,loewen,tarazona,roth,lutsko}
and recent density functional theory studies of wetting \cite{dhawan,berim,yatsyshin}.
We will start from the following grand potential functional:
\begin{eqnarray}
\label{03}
\Omega [\rho ({\rm {\bf r}})]
=
F_{\rm sr} [\rho ({\rm {\bf r}})]
+
F_{\rm lr} [\rho ({\rm {\bf r}})]
+
\int {\rm d {\bf r}}_1 \rho ({\rm {\bf r}_1})(V ({\rm {\bf r}_1})  - \mu),
\nonumber\\
F_{\rm sr} [\rho ({\rm {\bf r}})]
=
\int{\rm d {\bf r}}_1 
\rho({\rm {\bf r}}_1) 
\left(\ln\left(\Lambda^3\rho({\rm {\bf r}}_1)\right)
+\frac{-1+6v\rho({\rm {\bf r}}_1)-4v^2\rho^2({\rm {\bf r}}_1)}{\left(1-v\rho({\rm {\bf r}}_1)\right)^2}
\right),
\nonumber\\
F_{\rm lr} [\rho ({\rm {\bf r}})]
=
\frac{1}{2} \int_{\vert {\bf r}_1 - {\bf r}_2 \vert \ge \sigma }
{\rm d {\bf r}}_1 {\rm d{\bf r}}_2
\rho({\rm {\bf r}}_1) \rho({\rm {\bf r}}_2)
\left(
-\frac{3 a(\vert{\bf {E}}\vert) \sigma^3}{2\pi}\frac{1}{\vert {{\bf r}}_1 - {{\bf r}}_2 \vert^6}
\right),
\end{eqnarray}
which accounts for
the short-range repulsion $F_{\rm sr} [\rho ({\rm {\bf r}})]$ of hard-core spheres having the diameter $\sigma$
and
the long-range attraction $F_{\rm lr} [\rho ({\rm {\bf r}})]$, which depends on the external electric field.
Moreover,
$V({\bf r})$ is the external potential and $\mu$ is the chemical potential.
For $F_{\rm sr} [\rho ({\rm {\bf r}})]$ we use the local density approximation 
which would yield the Carnahan-Starling equation of state in the uniform limit,
see Appendix.
For $F_{\rm lr} [\rho ({\rm {\bf r}})]$ we use the mean-field approximation.
Such approximations completely neglect the local correlation structure around an atom
and more refined treatments are known for both contributions,
of the short-range repulsion \cite{tarazona,roth} and of the long-range attraction \cite{sokolowski,wadewitz}.
Nevertheless, the adopted treatment is suitable for the purposes of the present study
and more sophisticated approximations go beyond the scope of the present paper.

In what follows we also need to know the explicit form for the external potential $V({\bf r})$
which describes the interaction between the solid wall (substrate) and the atoms of fluid.
We assume that the solid wall, say, for $z<0$ is formed
with uniformly distributed two-level atoms with the density $\rho_s$,
which interact with the fluid two-level atoms via the same potential as in the fluid
(see, e.g., Ref.~\onlinecite{malijevsky}).
The long-range (i.e., $z\ge \sigma$) contribution of the semi-infinite planar solid wall
to $V_s(x,y,z)=V_s(z)$
is calculated by integrating the long-range interatomic interaction 
$-3a(\vert{\bf{E}}\vert)\sigma^3/(2\pi R^6)$
(cf. Eq.~(\ref{02}))
\begin{eqnarray}
\label{04}
V_s(z)
=
\rho_s
\int_{-\infty}^{\infty} {\rm d} x^\prime
\int_{-\infty}^{\infty} {\rm d} y^\prime
\int_{-\infty}^{0}      {\rm d} z^\prime
\left(-\frac{3a(\vert{\bf{E}}\vert)\sigma^3}{2\pi} 
\frac{1}{\sqrt{(x-x^\prime)^2 + (y-y^\prime)^2 + (z-z^\prime)^2}^6}\right)
=
- \frac{\rho_s a(\vert{\bf{E}}\vert) \sigma^3}{4} \frac{1}{z^3}.
\end{eqnarray}
In what follows we assume $\eta_s= \rho_s v=1$ for convenience.
Moreover, we set $V_s(z)=\infty$ for $0\le z<\sigma$.
Clearly, we have assumed that all three phases are influenced by the electric field.
Such a case is also experimentally realizable, 
see Ref.~\onlinecite{bateni},
where the used experimental setup was designed in such a way that the electric field was applied to all three interfaces.

The following remark about the elaborated theory is in order here.
As can be seen from the consideration above,
the electric field enters the theory only through the increase of the van der Waals interactions constant
which is simply multiplied by $1+2\aleph^2{\cal{E}}^2$.
This means that the electric field may be eliminated from the theory 
after introducing an appropriate energy unit.
For example,
after introducing the critical temperature $T_c(\vert{\bf{E}}\vert)$ as the energy unit
all dimensionless quantities should be already independent of the field.
Calculations reported in the next section confirm this observation.
Of course, this feature would be not present in more advanced consideration of the electric field.

The equation for the equilibrium density $\rho({\bf r})$ is given by
$\delta \Omega [\rho ({\rm {\bf r}})]/\delta \rho ({\rm {\bf r}}) =0$ \cite{evans,oxtoby-review,derzhko-myhal-lectures,loewen,tarazona,roth,lutsko}.
Substituting its solution into Eq.~(\ref{03}) 
we get the value of the grand potential of the nonuniform system under consideration $\Omega(T,\mu,{\cal{V}})$.
Here ${\cal{V}}$ is the volume of the system \cite{evans,oxtoby-review,derzhko-myhal-lectures,loewen,tarazona,roth,lutsko}.

\section{Bulk and surface properties}

Considering on the basis of Eq.~(\ref{03}) with $V({\rm {\bf r}})=0$ the bulk properties,
when $\rho({\bf r})=\rho$,
we find 
the critical density $\eta_c {\approx} 0.130\,44$,
the critical temperature $\tau_c({\cal {E}})\approx 0.001\,965\,18 \aleph^2(1+2\aleph^2{\cal {E}}^2)$,
and the critical pressure $\pi_c({\cal {E}})\approx 0.000\,092\,02 \aleph^2(1+2\aleph^2{\cal {E}}^2)$
of the fluid at hand;
here we have introduced the dimensionless variables
$\eta=\rho v$,
$\tau=T/(E_1-E_0)$,
$\pi=pv/(E_1-E_0)$,
see Ref.~\onlinecite{derzhko-myhal-zhfd}.
Within the adopted approach,
the critical density is independent of the field
but the critical temperature and the critical pressure increase by the factor $1+2\aleph^2{\cal {E}}^2$.
At temperatures below the critical temperature $T_c$
the fluid can be in the form of two coexisting phases (liquid and vapor).
In what follows we consider just such temperatures $T<T_c$.

Let us explain how to get the contact angle $\theta$.
First we calculate the liquid -- vapor surface tension $\gamma_{lv}$.
To this end,
we consider a nonuniform fluid at $T<T_c$ in the form of two phases in equilibrium with the planar interface
(see also Ref.~\onlinecite{derzhko-myhal-zhfd}).
For computation purposes, 
it is useful to assume 
that the fluid is within the cylindric vessel
of the radius ${\cal {R}}\rightarrow \infty$ and the height ${\cal {L}}$,
direct the $z$ axis of the coordinate system along the cylinder axis,
and take the origin of the coordinate system in the middle of the height.
We seek for the equilibrium density $\rho(z)=\rho(x=0,y=0,z)$
(i.e., along the cylindric axis)
and the equation for $\rho(z)$ 
after taking the limit ${\cal {R}}\rightarrow \infty$ 
has no traces of the adopted specific (cylindric) geometry.
Moreover,
we know the pressure $p(T)$ and the chemical potential $\mu(T)$ of the two-phase system at hand.
We put $V(z) = 0$, but seek for the solution for the equilibrium density $\rho(z)$ which depends on the height $z$.
The solution for the density profile $\rho(z)$
gives the value of the grand potential of the two-phase fluid in the cylinder
$\Omega(T,\mu(T),\pi{\cal{R}}^2{\cal{L}})$.
The surface tension follows from the relation
$\gamma_{lv}(T) =
(\Omega(T,\mu(T),\pi{\cal{R}}^2{\cal{L}})+p(T)\pi{\cal {R}}^2{\cal {L}})/(\pi{\cal {R}}^2)$.

The interface surface tensions substrate -- liquid  $\gamma_{sl}$ or substrate -- vapor  $\gamma_{sv}$ 
are calculated along the same lines,
however,
one has to take into account the potential of substrate  $V_s({\rm {\bf r}})$ (\ref{04}),
which is situated, say, at $z=0$.
We initialized the system in the cylindric vessel 
with the liquid density (i.e., $\rho (z)= \rho_l$) or the vapor density (i.e., $\rho (z)= \rho_v$)
if $z\gg 0$.
Then we find the equilibrium density $\rho(z)$, the grand potential $\Omega(T,\mu,\pi{\cal{R}}^2{\cal{L}}/2)$,
and, as a result, the values of $\gamma_{sl}(T)$ or $\gamma_{sv}(T)$.

Finally, the contact angle $\theta$ is defined by Young's equation
\begin{eqnarray}
\label{05}
\gamma_{s v} - \gamma_{s l} - \gamma_{l v}\cos\theta = 0.
\end{eqnarray}
Equation~(\ref{05}) completes the calculation of the contact angle $\theta(T,\vert{\bf{E}}\vert)$
starting from the interparticle interactions.

\begin{figure}
\center{\includegraphics[width=85mm]{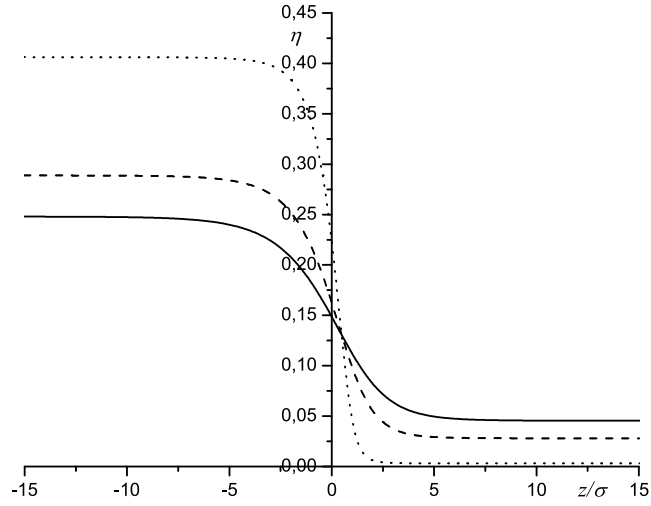}}
\caption[]{Liquid -- vapor density profile $\eta(z)$ at the temperature $\tau = 0.9\tau_c(0) \approx 0.001\,769$:
${\cal{E}}=0$ (solid),
${\cal{E}}=0.2$ (dashed),
and
${\cal{E}}=0.5$ (dotted).}
\label{fig03}
\end{figure}

\begin{figure}
\center
{\includegraphics[width=85mm]{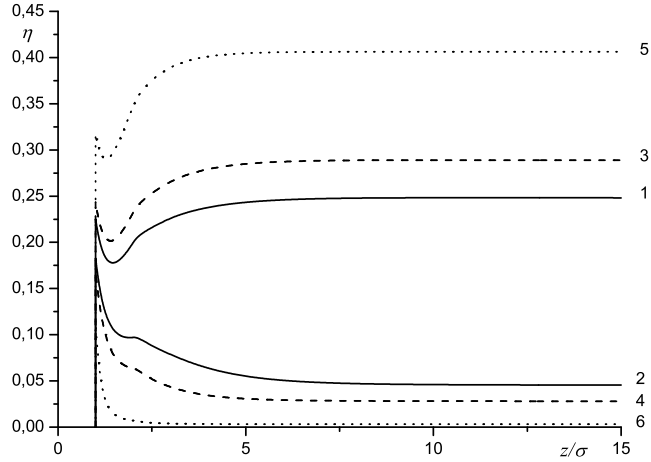}}
\vspace{3mm}
\caption[]{Density profile $\eta(z)$ near substrate (at $z=0$) at the temperature $\tau = 0.9\tau_c(0) \approx 0.001\,769$
without the electric field ${\cal{E}}=0$ (solid curves 1 and 2),
at ${\cal{E}}=0.2$ (dashed curves 3 and 4), 
and 
at ${\cal{E}}=0.5$ (dotted curves 5 and 6). 
Curves 1, 3, and 5 correspond to the case of liquid near substrate,
curves 2, 4, 6 correspond to the case of vapor near substrate.}
\label{fig04}
\end{figure}

Next, we turn to our findings.
Density profiles are shown in Figs.~\ref{fig03} and \ref{fig04}.
Some dependences of the dimensionless surface tensions $\Gamma =\gamma\sigma^2/(E_1-E_0)$ 
and those of the contact angle $\theta$
on the dimensionless value of the electric field strength ${\cal{E}}$
are reported in Table~\ref{tab1}.
The results in Figs.~\ref{fig03}, \ref{fig04} and Table~\ref{tab1}
refer to a particular representative value of the dimensionless temperature $\tau \approx 0.001\,769$
(this is $0.9\tau_c(0)$, 
where $\tau_c(0)$ denotes the dimensionless critical temperature without the field, i.e., at ${\cal{E}}=0$).
The results for the contact angle $\theta(\tau,{\cal{E}})$ obtained on the basis of Eq.~(\ref{05}) 
are reported in Figs.~\ref{fig01} and \ref{fig02}.
Bearing in mind a plausible experimental setup 
when the electric field is switching on at constant temperature,
we present all calculations at fixed $T$,
or more precisely, in the units proportional to $T_c({\bf{E}}=0)$, but not $T_c(\vert{\bf{E}}\vert)$.

\begin{table}
\begin{center}
\caption{Dependence of the dimensionless interface surface tensions
$\Gamma_{lv}$ (liquid -- vapor),
$\Gamma_{sl}$ (substrate -- liquid), 
$\Gamma_{sv}$ (substrate -- vapor),
and the contact angle $\theta$ (in degrees) 
on the dimensionless value of the electric field strength ${\cal{E}}$
at the temperature $\tau = 0.9\tau_c(0) \approx 0.001\,769$.
\label{tab1}}
\begin{tabular}[t]{|c|c|c|c|c|}
\hline
~ ${\cal{E}}$ ~ & ~ $\Gamma_{lv}$    ~ & ~ $\Gamma_{sl}$     ~ & ~ $\Gamma_{sv}$    ~ & ~ $\theta$   ~\\
\hline
 ~  0  ~        & ~ $0.000\,168\,9$  ~ & ~ $-0.000\,281\,8$  ~ & ~ $-0.000\,200\,9$ ~ & ~ $61^\circ$ ~\\
\hline
 ~  0.1 ~       & ~ $0.000\,219\,6$  ~ & ~ $-0.000\,281\,7$  ~ & ~ $-0.000\,195\,9$ ~ & ~ $67^\circ$ ~\\
\hline
 ~  0.2 ~       & ~ $0.000\,391\,2$  ~ & ~ $-0.000\,264\,9$  ~ & ~ $-0.000\,179\,9$ ~ & ~ $77^\circ$ ~\\
\hline
 ~  0.3 ~       & ~ $0.000\,725\,7$  ~ & ~ $-0.000\,193\,9$  ~ & ~ $-0.000\,151\,9$ ~ & ~ $87^\circ$ ~\\
\hline
 ~  0.4 ~       & ~ $0.001\,264\,6$ ~  & ~ $-0.000\,029\,5$  ~ & ~ $-0.000\,114\,7$ ~ & ~ $94^\circ$ ~\\
\hline
 ~  0.5 ~       & ~ $0.002\,041\,6$  ~ & ~ $~~0.000\,260\,7$ ~ & ~ $-0.000\,075\,3$ ~ & ~ $99^\circ$ ~\\
\hline
\end{tabular}
\end{center}
\end{table}

Density profiles in Fig.~\ref{fig03} show a diffused boundary between two phases, liquid and vapor,
which becomes sharper as ${\cal{E}}$ increases
(compare the solid and dotted curves).
This can be explained by an increase of the interatomic attraction as ${\cal{E}}>0$
which results in an increase of $T_c$ 
and therefore the fluid at the fixed temperature $\tau=0.9\tau_c(0)$ turns out to be farther from the critical region.
Some structure around the solid substrate seen in Fig.~\ref{fig04}
is due to the hard-core-sphere repulsion:
It manifests itself for $1\le z/\sigma\le 2$ \cite{balescu}.
It is better pronounced in the case of a more dense liquid phase
(curves 1, 3, 5)
and almost disappears for vapor
(curves 2, 4, 6).
Furthermore,
from Fig.~\ref{fig01} one concludes that the wetting temperature $T_w$ increases after the field has been switched on.
The temperature dependences of the contact angle in Fig.~\ref{fig01} satisfy the relation
$1-\cos\theta\propto (T_w-T)^{2-\alpha_s}$ with $\alpha_s=1$,
thus implying the wetting transitions of first order \cite{bonn-ross}.
From Fig.~\ref{fig02} one concludes that the contact angle grows with the increasing of the field strength.
For temperatures close to $T_w<T_c$ the change of $\theta$ is rather steep.
Moreover, 
$\theta$ may cross $90^{\circ}$
indicating that the hydrophilic surface ($\theta<90^\circ$) becomes hydrophobic ($\theta>90^\circ$).
For example,
for $\tau/\tau_c(0)=0.728\,4$ the contact angle crosses $90^\circ$ as ${\cal{E}}$ varies from $0.005$ to $0.010$. 
Clearly,
the actual value of $\theta$ follows from Eq.~(\ref{05})
and hence is determined by the interplay of surface tensions $\gamma_{lv}$, $\gamma_{sl}$, and $\gamma_{sv}$
at a given temperature and electric field strength magnitude.

\section{Discussion and conclusions}

Let us discuss the obtained results.
For the case of the considered two-level-atom fluid,
a nonzero electric field increases the long-range attraction in the system.
This leads to an increase of the critical temperature of the fluid
$T_c(\vert{\bf{E}}\vert)>T_c({\bf{E}}=0)$
and therefore,
after the field is switched on at constant temperature the two-phase state gets farther from the critical region.
All surface tensions increase with the increase of the field,
see Table~\ref{tab1}.
According to Eq.~(\ref{05}),
$\gamma_{lv}>0$ influences the value of $\cos\theta$ but not the change of its sign.
As can be seen from Table~\ref{tab1},
$\gamma_{sl}$ grows and changes its sign as the field increases.
As a result, 
$\cos\theta$ may change its sign 
and the hydrophilic surface ($\cos\theta>0$) change to the hydrophobic one ($\cos\theta<0$).
We adopted a very simple model for the substrate.
The external potential $V_s({\bf{r}})$ representing the substrate may be made smaller 
(e.g., by a decrease of $\eta_s$).
Then the role of the substrate diminishes:
It behaves as a hydrophobic surface even in the absence of the field
and is less sensitive to the presence of the field.
However, qualitatively the effect of the field remains the same: 
The contact angle grows with the increase of the field.
Finally it is worth noting, 
that the growth of $\theta$ slows down for relatively large fields, see Fig.~\ref{fig02}, 
and our consideration (which is valid for small ${\cal{E}}$ only)
does not give hints for complete drying induced by the field.

It is also in order to make here a remark concerning the electric-field-strength scale.
This scale is defined by
$\vert{\bf{E}}_0\vert\equiv\vert{\bf{p}}\vert/r_0^3$
and is of the order of $10^{10}$ volts per meter.
Such large values of $\vert{\bf{E}}_0\vert$ may be expected, since we deal with atomic-scale electric fields.
However,
if the temperature is very close to (just below) the wetting temperature $T_w<T_c$
even small electric field strengths can produce noticeable changes in $\theta$.

According to our study,
the treatment of the electric field effects on the basis of the Lennard-Jones fluid  
(see, e.g., Refs.~\onlinecite{dhawan,sartarelli})
should imply a change of the Lennard-Jones potential parameters
to be in agreement with the increase of the van der Waals interactions constant by $1+2\aleph^2{\cal{E}}^2$.
Finally,
the elaborated scheme can be also applied to examine the wetting in the presence of excited atoms
which may appear as a result of resonance irradiation \cite{yukhnovskii}.

A few words about a comparison with the outcomes of alternative approaches 
which permits to test the quality of the obtained results 
are in order here.
The most important test requires ab initio calculations of the effective interatomic interactions
in the presence of the electric field
since this information would check the dependence of the initial grand potential functional (\ref{03}) on the field.
However, such simulations are far beyond the scope of the present study.
On the other hand, 
the quality of the classical density functional theory results based on the simple grand potential functional (\ref{03})
for a system with the hard-core repulsion and the van der Waals attraction 
is known from previous studies (e.g., Ref.~\onlinecite{evans}).

To summarize,
we applied a classical density functional theory 
to a simple two-level-atom fluid 
to examine the effect of an external electric field on the wetting properties.
In the considered model the electric field couples to the transition electric dipole moment of atoms
resulting in the increase of the long-range interatomic attraction in the system.
Just below the wetting temperature
the electric field can increase noticeably the contact angle and lead to a passage from hydrophilicity to hydrophobicity.
Our calculations may refer to the noble fluids in a strong electric field.

\section*{Acknowledgments}

O.~D. acknowledges the kind hospitality of Prof.~Jozef Stre\v{c}ka 
(P.~J.~\v{S}af\'{a}rik University, Ko\v{s}ice, Slovakia)
during the CSMAG'16 conference in June of 2016.
O.~D. is grateful to the Organizing Committee of the 26th IUPAP International Conference on Statistical Physics
(Lyon, July 18-22, 2016)
for a financial support for attending the Conference.
O.~D. would also like to thank the Abdus Salam International Centre for Theoretical Physics (Trieste, Italy)
for partial support of this study through the Senior Associate Award.

\onecolumngrid

\section*{Appendix: The second virial coefficient of the two-level-atom fluid (\ref{02})}
\renewcommand{\theequation}{A\arabic{equation}}
\setcounter{equation}{0}

For the sake of being self-contained,
in this appendix, 
we provide some details necessary to understand the initial grand potential functional,
see Eq.~(\ref{03}).

We consider $N$ two-level atoms at sufficiently large interatomic distances 
$\vert {\bf {R}}_{ij}\vert=\vert {\bf {R}}_{i}-{\bf {R}}_{j}\vert$,
adopt the dipole approximation 
and use a convenient spin-1/2 representation \cite{excitons}
to write the electron subsystem Hamiltonian as
\begin{eqnarray}
\label{a01}
H({\bf{R}}_1,\ldots,{\bf{R}}_N)
=
\frac{N}{2}\left(E_0+E_1\right)+\left(E_1-E_0\right)\sum_{i=1}^Ns_i^z
+\frac{1}{2}\sum_{i,j=1 (i\ne j)}^N C_{ij}s_i^xs_j^x
+\sum_{i=1}^NB_is_i^x,
\end{eqnarray}
where $E_0$ and $E_1$ are the energies of the ground and excited states,
\begin{eqnarray}
\label{a02}
C_{ij}\equiv 4\alpha_{ij}\left(E_1-E_0\right)
=4\frac{\vert{\bf {p}}_{i}\vert \vert{\bf{p}}_{j}\vert}{\vert{\bf{R}}_{ij}\vert^3}\Phi_{ij},
\nonumber\\
\Phi_{ij}
=
\sin\theta_{{\bf{p}}_i}\sin\theta_{{\bf{p}}_j}\cos\left(\phi_{{\bf{p}}_i}-\phi_{{\bf{p}}_j}\right)
+
\cos\theta_{{\bf{p}}_i}\cos\theta_{{\bf{p}}_j}
\nonumber\\
-3
\left(
\sin\theta_{{\bf{p}}_i}\sin\theta_{{\bf{n}}_{ij}} \cos\left(\phi_{{\bf{p}}_i}-\phi_{{\bf{n}}_{ij}}\right)
+
\cos\theta_{{\bf{p}}_i}\cos\theta_{{\bf{n}}_{ij}}
\right)
\left(
\sin\theta_{{\bf{p}}_j}\sin\theta_{{\bf{n}}_{ij}} \cos\left(\phi_{{\bf{p}}_j}-\phi_{{\bf{n}}_{ij}}\right)
+
\cos\theta_{{\bf{p}}_j}\cos\theta_{{\bf{n}}_{ij}}
\right),
\end{eqnarray}
\begin{eqnarray}
\label{a03}
B_{i}\equiv \gamma_{i}\left(E_1-E_0\right)
=-2\vert{\bf p}_{i}\vert \vert{\bf E}\vert\chi_{i},
\nonumber\\
\chi_{i}
=
\sin\theta_{{\bf{p}}_i}\sin\theta_{{\bf{E}}}\cos\left(\phi_{{\bf{p}}_i}-\phi_{{\bf{E}}}\right)
+
\cos\theta_{{\bf{p}}_i}\cos\theta_{{\bf{E}}},
\end{eqnarray}
$\theta_{{\bf{p}}_i}$,
$\phi_{{\bf{p}}_i}$,
$\theta_{{\bf{n}}_{ij}}$,
$\phi_{{\bf{n}}_{ij}}$,
$\theta_{{\bf{E}}}$,
$\phi_{{\bf{E}}}$
are the angles that determine the orientation 
of the transition electric dipole moment of the $i$-th atom ${\bf{p}}_i$,
the unit vector ${\bf{n}}_{ij}={\bf{R}}_{ij}/\vert {\bf{R}}_{ij} \vert$,
and
the electric field ${\bf{E}}$.
The first two terms in Eq.~(\ref{a01}) describe a system of noninteracting two-level atoms,
the third one represents the dipole-dipole interaction between them,
and the last one corresponds to the interaction with the field.
To find the effective long-range interactions,
one has to calculate the eigenvalues of the Hamiltonian given in Eq.~(\ref{a01}).
Although this calculation is straightforward within the used spin-1/2 representation for not too large $N$,
in what follows we are interested in the case of small fields,
and therefore we may use the standard quantum-mechanical perturbation theory
assuming the interaction with the field to be small, 
i.e., $\gamma_i\ll 1$.
A correction to the ground-state energy of a single atom 
($N=1$, the third term in Eq.~(\ref{a01}) drops out)
appears in the second order and is given by the formula for $E_{N=1}$ in Eq.~(\ref{01}). 
For $N=2$ it is reasonable to assume in addition that $\alpha_{12}\ll 1$
(after such an assumption one gets the usual van der Waals interactions for ${\bf{E}}=0$)
and the second-order correction to the ground-state energy of two atoms is given by the formula for $E_{N=2}$ in Eq.~(\ref{01}).

Let us turn to statistical mechanics.
Presenting the grand partition function in the exponential form,
\begin{eqnarray}
\label{a04}
\Xi\equiv \sum_{N=0}^{\infty}z^N Z_N = \exp\left({\cal{V}}\sum_{l=1}^{\infty}z^l b_l\right),
\nonumber\\
{\cal{V}}b_1=Z_1,
\;\;\;
{\cal{V}}b_2=Z_2-\frac{1}{2}Z_1^2,
\;\;\;
\ldots,
\end{eqnarray}
where $z$ is the activity and ${\cal{V}}$ is the volume of the system,
we obtain the cluster expansion for the grand potential
\begin{eqnarray}
\label{a05}
-\frac{\Omega}{T{\cal{V}}}=zb_1+z^2b_2+\ldots,
\end{eqnarray}
which results in the virial equation of state
\begin{eqnarray}
\label{a06}
\frac{p}{T}=\rho+B_2\rho^2+\ldots,
\;\;\;
B_2=-\frac{b_2}{b_1^2},
\end{eqnarray}
where $\rho$ denotes the density of the system.
For the required canonical partition functions one has
\begin{eqnarray}
\label{a07}
Z_1=\frac{{\cal{V}}}{\Lambda^3}\left\langle \exp\left(-\frac{E_{N=1}}{T}\right) \right\rangle,
\nonumber\\
Z_2=\frac{{\cal{V}}}{2\Lambda^6}
\int_{\vert {\bf{R}}_{12}\vert \ge \sigma}{\rm{d}}{\bf{R}}_{12} \left\langle \exp\left(-\frac{E_{N=2}}{T}\right) \right\rangle,
\end{eqnarray}
where $\Lambda$ stands for the thermal de Broglie wavelength,
$E_{N=1}$ and $E_{N=2}$ are defined in Eqs.~(\ref{01}), (\ref{a02}), (\ref{a03}),
and the angle brackets mean the average over the orientations of transition dipole moments
\begin{eqnarray}
\label{a08}
\langle(\ldots)\rangle
=
\int{\rm{d}}\Omega_{{\bf{p}}_1}\ldots\int{\rm{d}}\Omega_{{\bf{p}}_N}(\ldots),
\;\;\;
\int{\rm{d}}\Omega_{{\bf{p}}_i}
=
\frac{1}{4\pi}\int_0^{2\pi}{\rm{d}}\phi_{{\bf{p}}_i} \int_0^{\pi}{\rm{d}}\theta_{{\bf{p}}_i}\sin\theta_{{\bf{p}}_i}. 
\end{eqnarray}
Bearing in mind that we assume the field to be small,
the orientational average (\ref{a08}) can be done using cumulant expansion
$\langle\exp x\rangle=\exp \left(\langle x\rangle +\left(\langle x^2\rangle -\langle x\rangle^2\right)/2+\ldots \right)$.
Keeping the terms up to ${\cal{O}}({\bf{E}}^2)$ only, we would need the following averages:
\begin{eqnarray}
\label{a09}
\langle\chi_i^2\rangle=\frac{1}{3},
\;\;\;
\langle\Phi_{12}^2\rangle=\frac{2}{3},
\;\;\;
\langle\chi_1\chi_2\Phi_{12}\rangle=\frac{1}{9}\left(1-\cos^2\theta_{{\bf{n}}_{12}}\right),
\;\;\;
\langle\chi_i^2\Phi_{12}\rangle=\frac{1}{45}\left(8+6\cos^2\theta_{{\bf{n}}_{12}}\right).
\end{eqnarray}
Equations~(\ref{a04}), (\ref{a07}), (\ref{a09}) give the explicit result for $b_1$ \cite{derzhko-myhal-zhfd}
and the formula for $b_2$ as a two-fold integral \cite{derzhko-myhal-zhfd}
which besides the integration over $R=\vert {\bf{R}}_{12}\vert$ contains the integration over $\theta_{{\bf{R}}_{12}}$, 
see Eq.~(\ref{a07}).
Introducing the variable $y=\cos\theta_{{\bf{R}}_{12}}$
one can do the integration over $y$ again with the help of the cumulant expansion 
with the accuracy up to the terms ${\cal{O}}({\bf{E}}^2)$.
The obtained cluster integrals give for the second virial coefficient $B_2$ in Eq.~(\ref{a06}) 
the formula for $B_2(T, \vert{\bf {E}}\vert)$ in Eq.~(\ref{02}).

Next, instead of the virial equation of state (\ref{a06}), (\ref{02}) we introduce an extrapolated equation of state
\begin{eqnarray}
\label{a10}
\frac{p}{T}
=\rho\frac{1+v\rho+v^2\rho^2-v^3\rho^3}{\left(1-v\rho\right)^3}-\rho^2\frac{a(\vert{\bf{E}}\vert)}{T}
\end{eqnarray}
and treating Eq.~(\ref{a10}) as an input after some simple standard assumptions arrive at the initial grand potential functional given in Eq.~(\ref{03}).


\begin{thebibliography}{99}

\bibitem{gennes}
P.~G.~de~Gennes,
Rev. Mod. Phys. {\bf 57}, 827 (1985).

\bibitem{armr}
M.~Rauscher and S.~Dietrich,
Annu. Rev. Mater. Res. {\bf 38}, 143 (2008).

\bibitem{bonn}
D.~Bonn, J.~Eggers, J.~Indekeu, J.~Meunier, and E.~Rolley,
Rev. Mod. Phys. {\bf 81}, 739 (2009).

\bibitem{saam}
W.~F.~Saam,
J. Low Temp. Phys. {\bf 157}, 77 (2009).

\bibitem{evans}
R.~Evans, 
Adv. Phys. {\bf 28}, 143 (1979);
R.~Evans,
Density Functionals in the Theory of Nonuniform Fluids.
{\it Fundamentals of Inhomogeneous Fluids}, edited by D.~Henderson
(Marcel Dekker, Inc., New York, 1992), pp.~85-175;
R.~Evans,
Density Functional Theory for Inhomogeneous Fluids I: Simple Fluids in Equilibrium.
{\it Lecture Notes at 3rd Warsaw School of Statistical Physics, Kazimierz Dolny, 27 June -- 3 July 2009}
(Warsaw University Press, 2010), p.~43.

\bibitem{oxtoby-review}
D. W. Oxtoby,
J. Phys.: Condens. Matter {\bf 4}, 7627 (1992).

\bibitem{derzhko-myhal-lectures}
O.~V.~Derzhko and V.~M.~Myhal,
{\it Selected topics on the theory of nonuniform classical fluids: A course of lectures}
(L'viv University, L'viv, 1999)
(in Ukrainian).

\bibitem{loewen}
H.~L\"{o}wen,
J. Phys.: Condens. Matter {\bf 14}, 11897 (2002).

\bibitem{tarazona}
P.~Tarazona, J.~A.~Cuesta, and Y.~Martinez-Rat\'{o}n, 
Density Functional Theories of Hard Particle Systems.
{\it Theory and Simulation of Hard-Sphere Fluids and Related Systems, 
Lecture Notes in Physics, Vol.~753},
edited by A.~Mulero
(Springer-Verlag, Berlin, Heidelberg, 2008), pp.~247-341.

\bibitem{roth}
R.~Roth,
J. Phys.: Condens. Matter {\bf 22}, 063102 (2010).

\bibitem{lutsko}
J.~F.~Lutsko,
Advances in Chemical Physics {\bf 144}, 1 (2010).

\bibitem{kang}
K.~H.~Kang,
Langmuir {\bf 18}, 10318 (2002).

\bibitem{bier}
M.~Bier and I.~Ibagon,
Phys. Rev. E {\bf 89}, 042409 (2014).

\bibitem{rui}
Z.~Rui, L.~Qi-Chao, W.~Ping, and L.~Zhong-Cheng,
Chin. Phys. B {\bf 24}, 086801 (2015).

\bibitem{bateni}
A.~Bateni, S.~Laughton, H.~Tavana, S.~S.~Susnar, A.~Amirfazli, and A.~W.~Neumann,
J. Colloid Interface Sci. {\bf 283}, 215 (2005).

\bibitem{mugele}
F.~Mugele and J.-C.~Baret,
J. Phys.: Condens. Matter {\bf 17}, R705 (2005).

\bibitem{vancauwenberghe}
V.~Vancauwenberghe, P.~Di~Marco, and D.~Brutin,
Colloids and Surfaces A: Physiochem. Eng. Aspects {\bf 432}, 50 (2013).

\bibitem{derzhko-myhal-zhfd}
O.~V.~Derzhko and V.~M.~Myhal,
J. Phys. Studies (L'viv) {\bf 1}, 402 (1997) 
(in Ukrainian).

\bibitem{dhawan}
S.~Dhawan, M.~E.~Reimel, L.~E.~Scriven, and H.~T.~Davis,
J. Chem. Phys. {\bf 94}, 4479 (1991).

\bibitem{berim}
G.~O.~Berim and E.~Ruckenstein,
J. Chem. Phys. {\bf 125}, 164717 (2006).

\bibitem{yatsyshin}
P.~Yatsyshin, N.~Savva, and S.~Kalliadasis,
J. Chem. Phys. {\bf 136}, 124113 (2012);
P.~Yatsyshin, N.~Savva, and S.~Kalliadasis,
J. Chem. Phys. {\bf 142}, 034708 (2015);
P.~Yatsyshin, N.~Savva, and S.~Kalliadasis,
J. Phys.: Condens. Matter {\bf 27}, 275104 (2015).

\bibitem{sokolowski}
S.~Sokolowski and J.~Fischer,
J. Chem. Phys. {\bf 96}, 5441 (1992).

\bibitem{wadewitz}
T.~Wadewitz and J.~Winkelmann,
J. Chem. Phys. {\bf 113}, 2447 (2000).

\bibitem{malijevsky}
A.~Malijevsk\'{y} and A.~O.~Parry, 
J. Phys.: Condens. Matter {\bf 25}, 305005 (2013);
A.~Malijevsk\'{y}, 
J. Phys.: Condens. Matter {\bf 25}, 445006 (2013);
A.~Malijevsk\'{y}, 
J. Chem. Phys. {\bf 141}, 184703 (2014).
If one assumes that the solid wall is formed by uniformly distributed two-level atoms 
placed within the plane $z=0$,
we arrive at $1/z^4$ decay law for $V_s(z)$ (instead of $1/z^3$ for $V_s(z)$ in Eq.~(\ref{04})).

\bibitem{balescu}
R.~Balescu,
{\it Equilibrium and Nonequilibrium Statistical Mechanics}
(Wiley-Interscience, New York, 1975).

\bibitem{bonn-ross}
D.~Bonn and D.~Ross, 
Rep. Prog. Phys. {\bf 64}, 1085 (2001).

\bibitem{sartarelli}
S.~A.~Sartarelli and L.~Szybisz,
Phys. Rev. E {\bf 80}, 052602 (2009).

\bibitem{yukhnovskii}
I.~R.~Yukhnovskii, O.~V.~Derzhko, and R.~R.~Levitskii,
Physica A {\bf 203}, 381 (1994);
O.~Derzhko, R.~Levitskii, and O.~Chernyavskii,
Condens. Matter Phys. (L'viv) {\bf 6}, 35 (1995).

\bibitem{excitons}
V.~M.~Agranovich,
{\it Theory of Excitons}
(Nauka, Moscow, 1968)
(in Russian);
S.~Takeno and M.~Mabuchi,
Progress of Theoretical Physics {\bf 50}, 1848 (1973);
I.~R.~Yukhnovskii, R.~M.~Kadobyanskii, R.~R.~Levitskii, and O.~V.~Derzhko,
Ukr. Fiz. Zh. {\bf 34}, 300 (1989)
(in Russian).

\end{thebibliography}
\end{document}